\documentstyle[aaspp4]{article}
\def\ea{{\it et al.} }
\def\ref{\par\noindent\hangindent 20 pt}

\begin{document} 

\input{psfig.tex}

\title{\bf On The Parent Population of Radio Galaxies and the FR~I -- FR~II
Dichotomy}

\author{Riccardo Scarpa }
\affil{European Southern Observatory}
\author{C. Megan Urry}
\affil{Space Telescope Science Institute}
\authoraddr{European Southern Observatory, 3107 Alonso de Cordova, Santiago, Chile}
\authoraddr{Space Telescope Science Institute, 3700 San Martin Dr., 
Baltimore, MD 21218, USA}
\authoremail{rscarpa@eso.org}

\begin{abstract}
We test the hypothesis that radio galaxies are a random subset of
otherwise normal elliptical galaxies.  Starting with the observed
optical luminosity functions for elliptical galaxies, we show that the
probability of an elliptical forming a radio source is a continuous,
increasing function of optical luminosity, proportional to $L^{2\pm0.4}$.
With this probability function and the luminosity function of normal
elliptical galaxies as input to Monte Carlo simulations, we reproduce
the observed distribution of radio galaxies in the radio-optical
luminosity plane. Our results show that radio galaxies are a
luminosity-biased but otherwise random sample of elliptical galaxies.
This unified view of radio-loud and radio-quiet ellipticals also
explains the well known difference of $\sim 0.5$~mag in average
optical luminosity between FR~I and FR~II radio galaxies as a simple
selection effect. Specifically, FR~II appear preferentially in smaller
galaxies because both radio and optical luminosity functions are
steep, so there is a negligible probability of observing a powerful
radio source associated with a bright galaxy; no intrinsic physical
differences between FR~I and FR~II host galaxies are required.  
\end{abstract} 

\section{Introduction}

Some elliptical galaxies are radio loud --- i.e., they have
bright nuclear radio sources emitting $>10^{22}$~W/Hz at 1.5~GHz 
 ---  but most are not. Understanding the reason for this dichotomy is
extremely important for fully understanding elliptical galaxies.
Currently the most promising explanation for the nuclear activity is
the presence of gas accretion around a massive black hole.
Then either a small fraction of ellipticals have such massive objects in 
their center, or all have black holes but the phase of activity is
short compared with galaxies lifetime.

There is growing evidence that massive black holes are present 
in the center of most (perhaps all) elliptical galaxies 
(Ho 1998; Richstone \ea 1998; van der Marel 1999; Macchetto 1999). 
In particular
a tight correlation between bulge and black hole mass has been
reported (Ferrarese \& Merritt 2000; Gebhardt et al. 2000; Magorrian \ea 1998).
This suggests all elliptical galaxies have the potential to become
active and only those currently accreting are identified
as active galactic nuclei (AGN). 

It is in fact difficult to distinguish radio-loud and 
radio-quiet ellipticals solely from their optical properties. 
Regardless of the level of nuclear 
activity, elliptical galaxies occupy the same fundamental plane, have similar 
ellipticity, isophotal twists, and colors distributions 
(Homabe \& Kormendy 1987; Ledlow \& Owen 1995; Govoni \ea 2000). 
These results can be extended to the most active 
radio-loud AGN (meaning blazars, BL Lac objects, and
intermediate quasars; Falomo et al. 2000; Scarpa et al. 2000; Urry \ea 2000).

There is also no clear connection between nuclear activity and 
the Megaparsec-scale cluster environment (Fanti 1984; Ledlow \& Owen 1995),
and the importance of galaxy interactions remains unclear. In particular
some interacting galaxies are not associated with powerful radio sources and,
conversely, some radio sources are associated with isolated galaxies.
Thus, galaxy interaction may not be necessary or sufficient
for transforming an elliptical galaxy into a radio galaxy, although
in the former case the radio activity may not have started yet, 
and in the latter case, if a small galaxy were accreted 
the global structure of the galaxy might not have been altered significantly. 
Whatever the case, a necessary condition for making a radio galaxy
seems to be the presence of a massive black hole, 
which seems to be a common feature of all galaxies.

There has been speculation that the central black holes 
of radio-loud and radio-quiet AGN differ in some quantized way. 
For example, perhaps black hole spin determines whether an AGN
forms a radio source (with spin) or does not (Rees 1997; 
Wilson \& Colbert 1995).
Yet recent work argues against this standard view of radio sources 
as ``on" (radio-loud) or ``off" (radio-quiet).
Indeed radio-selected quasars occur at all levels of radio-loudness,
with no apparent bimodality (White et al. 2000).

In this paper we test the possibility that all elliptical galaxies can
host radio sources of any power or radio class.  We use Monte Carlo
simulations to see whether observed samples of radio galaxies can be
random selections of elliptical galaxies; that is, whether it is
possible to find a direct link between the optical luminosity
functions of radio-quiet and radio-loud galaxies.  A somewhat similar
approach to unifying AGN was first discussed by Rowan-Robinson (1977),
although at that time there was less information about the optical
luminosity function of ellipticals than we have now.

We start by noting that ellipticals of different size might well 
have different probabilities of forming strong radio sources,
as was first deduced by Rowan-Robinson. The bright end
of the optical luminosity function (LF) is so steep that for every galaxy with
$M_R=-25$~mag, there are $\sim 10^{3}$ galaxies with $M_R=-22$~mag,
yet in complete samples of radio sources (Ledlow \& Owen 1996; Govoni
et al. 2000) a roughly constant number of radio galaxies is observed
between $-25<M_R<-21$~mag. This indicates that the probability of
observing radio emission from a galaxy increases strongly with its
optical luminosity, $L$. The LFs of radio-quiet and radio-loud galaxies
then have to be linked by a probability function, $S(L)$, whose functional
form can be derived from observations.

The goal of this work is to reproduce the optical LF of radio galaxies only,
not to describe their radio properties.
Nevertheless, making additional assumptions about the 
radio LF and the division between FR I and FR II galaxies, 
the simulated distribution of radio galaxies in the radio-optical 
luminosity plane compares well with actual radio galaxy samples.
A result that will be used to propose an explanation for the well known
difference in optical luminosity between 
FR~I and FR~II radio galaxies (Fanaroff \& Riley 1974); FR~I being
systematically brighter than FR~II by $\sim 0.5$ magnitudes 
(Owen \& Laing 1989; Ledlow \& Owen 1996; Smith \& Heckman 1989; 
Govoni \ea 2000). 

We outline the basic assumptions and calculations in the next section 
and derive the probability function  $S(L)$ in \S~3. Then we compare 
the predictions of this scenario with the observed data in
\S~4, and summarize our conclusions in \S~5. 

\section{Basic Assumptions}

Using a statistical (Monte Carlo) approach, we investigate the possibility
that all elliptical galaxies may host a radio source of arbitrary power. 
We start with the following general assumptions based on
the available empirical results for radio galaxies.

\begin{enumerate}

\item
The distribution of elliptical galaxies in
optical luminosity, $L$, is given by a Schechter luminosity function:

\begin{equation}
\Phi(L) = \frac{\Phi ^*}{L^*} ~(\frac{L}{L^*})^\alpha~ e^{-(\frac{L}{L^*})} ~,
\end{equation}

where $\alpha$ sets the slope of the LF at the faint end, 
$L^*$ gives the characteristic luminosity above which the number of galaxies 
falls sharply, and $\Phi^*$ (in units of galaxies Mpc$^{-3}~L^{-1}$)
sets the overall normalization of the galaxy density. 

\item
All elliptical galaxies of all optical luminosities 
have the potential to be radio sources, with a probability 
$S(L)$ proportional to some power $h$ of their optical luminosity:
$S(L)= S^* (\frac{L}{L^*})^h$. $S^*$ sets the overall normalization
of the function, and $S(L)$ is dimensionless.

\item
Regardless of their optical luminosity, once 
activated, all ellipticals produce radio sources with the same 
power-law distribution $N(P)\propto P^{\beta}$ (in units of $P^{-1}$).
Note that only the shape of the radio LF is used, the
normalization being set automatically by the optical LF.

\item
In the radio-optical luminosity plane, FR~I and FR~II are separated by
a transition line roughly proportional to $L^2$, 
with normalization depending on the frequency under consideration. 
Here we use $\log P = 1.7 \log L + 9.83$ 
(following Ledlow \& Owen 1996; normalized for R-band luminosity).

\end{enumerate}

An implicit assumption is that the active phase lasts on average
the same time in all ellipticals; otherwise, we would have to include
an explicit dependence of duration or duty-cycle on some property of the
elliptical galaxies.
Since the activity phase is driven by the availability of gas 
close to the nucleus rather than the global size of the galaxy, 
we believe this implicit assumption is reasonable.

 From points (1) and (2), 
the number of radio galaxies with optical luminosity $L$ is the product
of the optical LF times the probability $S$:

\begin{equation}
N(L) = \Phi(L) ~ S(L) = \frac{\Phi ^*}{L^*}~ S^* (\frac{L}{L^*})^{(\alpha+h)}
~ e^{-(\frac{L}{L^*})} ~,
\end{equation}

\noindent
which is a Schechter function with exponent $(\alpha + h)$. From 
the bivariate LF we know that approximately $S(L)\propto L^2$, that is $h\sim2$
(Ledlow \& Owen 1996). Because of this strong dependence on $L$, 
$N(L)$ is sharply peaked around $M\sim -24$~mag (Fig.~1),
so the number of radio sources in dim galaxies is very small 
despite the huge number of such galaxies.
This is the effect of $S(L)$; without such a steep dependence on
optical luminosity, the vast majority of radio 
sources would be associated with small/faint optical galaxies.

Assumptions (1) and (2) alone are already sufficient
to determine the value of $h$ necessary to link the luminosity
functions of radio-quiet and radio-loud galaxies (see next section).
We can go further, however, and test the model against
known properties of radio galaxies (hence assumptions 3 and 4).
The final distribution of radio sources in the radio-optical 
luminosity plane is given by the product of $N(L)$ times the function 
$N(P)$, which is assumed to be the same for all optical luminosities (point 3).

Once a random set of values has been extracted from equation (2),
these ``radio galaxies'' are divided into FR~I and FR~II according 
to point (4). No specific physical meaning is
assigned to the two types of radio sources. In particular,
the slope of the dividing line is not related to the slope of the 
probability function $S(L)$; we simply use it to divide
the two classes so that the average optical luminosities
of each can be derived. 

To actually compare predicted and observed counts of radio sources, it
is necessary to fold in the different volumes over which a source of a
given radio flux can be observed. Most surveys are flux
limited, so that bright sources are preferentially discovered. 
For radio sources, the limiting volume comes usually from the radio flux.
If $f_{lim}$ is the radio flux limit of the survey at a give 
frequency, and $P$ the power of a radio source at the same frequency, 
then this radio source could be observed out to a luminosity distance of:
\begin{equation}
d_L=\sqrt{\frac{P(1+z)^{1+s}}{4\pi f_{lim}}} ~,
\end{equation}
from which we obtain the volume $V(P)$ over which the sources of 
power $P$ can be discovered. The K correction, $(1+z)^{1+s}$,
appropriate to a power-law spectrum $F(\nu)\propto \nu^s$, 
has been also included\footnote{Due to the stochastic nature of the 
Monte Carlo simulation, the modest flux correction due to the K correction 
has only a minor effect, if any, on our result.}. 
At low redshift the volume increases as $P^{3/2}$, strongly 
favoring powerful sources. At higher redshift $V$ increases less rapidly
than in the Euclidean case, depending on the adopted cosmology.

Combining all previous equations, the total number of FR~I galaxies 
expected in a flux limited survey is:
\begin{equation}
N_{FR~I}= \int_{0}^{\infty}{ 
 \int_{0}^{P_{transition}}{\Phi(L) ~S(L) ~N(P) ~V(P) ~dL ~dP} ~,
}
\end{equation}
where $P_{transition}$ is the transition power from FR~I to FR~II
(itself a function of $L$). A similar expression with integration limits
from $P_{transition}$ to $\infty$ gives the number of FR~II galaxies.

In the previous equations there is only one free parameter,
the exponent describing the probability of observing radio
emission from an elliptical galaxy, $h$,
all other quantities being taken from the literature.
The slope of the radio luminosity function 
(Auriemma \ea 1977; Toffolatti \ea 1987; Urry \& Padovani 1995; 
Ledlow \& Owen 1996) is well defined, $\beta = -2$
(for $N(P)$ in units of Mpc$^{-3} P^{-1}$; it would be 
$N(P)\propto P^{-1}$ in units of Mpc$^{-3}/\Delta \log P$). 
For the optical LF we use $M_R^*=-22.8$~mag 
(in the Cousins R band; H$_0=50$ km/s/Mpc; q$_0=0$) and $\alpha=+0.2$, 
as found for elliptical galaxies in the
Stromlo-APM experiment (Loveday \ea 1992).
Other authors have found slightly different values for $M^*$ and $\alpha$
(Muriel \ea 1995; Lin \ea 1996; Zucca \ea 1997), 
and there is also some freedom in the location of the FR~I--II 
transition line. We will later investigate the effect of changing these values. 

\section{Constraining the Probability Function $S(L)$}

The normalized cumulative distribution of radio galaxies 
in optical luminosity is obtained by integrating equation (2)
(which is based only on assumptions 1 and 2):
it is an incomplete gamma function, $\gamma(1+\alpha+h , L/L^*)$. 
Both constants $\Phi^*$ and $S^*$ cancel out, 
leaving $h$ as the only free parameter, which can be derived from
fitting  observed optical magnitudes for radio galaxies.

We collect three well-studied samples of radio galaxies to which we
can fit this distribution (see Fig.~2).  However, the three observed
distributions are not internally consistent: according to a K-S test,
the probability that they are drawn from the same parent population is
only $\sim 5$\%.  Therefore, no one model (or one value of $h$) can
explain all three simultaneously.  There are many reasons why these
samples may differ, including the different photometric systems used;
magnitudes integrated over different apertures, with different
assumptions about the light distributions; different extinction and K
corrections applied; and different telescopes/instruments used.
Therefore, to obtain a robust estimate of the true optical luminosity
distribution of radio galaxies, we combined the data from all three
works.  This combined distribution is plotted in Figure~3, together
with the incomplete gamma function $\gamma(1+\alpha+h, L/L^*)$ for
$h=2\pm0.4$, determined using a maximum likelihood approach.  Fitting
the three samples of radio galaxies individually yields values for $h$
within this interval.

Changing $L^*$ or $\alpha$ in the optical LF for elliptical galaxies
will force a change in $h$. The sum $h+\alpha$ enters in the gamma
function, so any variation of $\alpha$ can be compensated by an equal
and opposite variation of $h$, without affecting any other results
presented here.  Changing $L^*$, or equivalently $M^*$, has the effect
of changing the slope of the theoretical cumulative distribution. A
brighter $M^*$ gives a steeper distribution (compared to the one in
Fig.~3) which would actually agree better with observations. For
$M^*=-23$~mag, the best fit is obtained for $h=1.6$.  For fainter
$M^*$ the distribution becames flatter and the statistical agreement
with the data becomes worse.  For $M^*=-22.6$~mag we get $h=2.5$.  To
the extent that the luminosity function of elliptical galaxies is
uncertain, we can not know $h$ precisely.

\section{Comparison with Observed Radio Samples}

Having fixed $h$, we now make use of assumptions (3) and (4) 
and Equation~4 to populate the radio-optical luminosity
plane, and to compare our result with data from flux-limited radio surveys.

\subsection{An Illustrative Comparison with a Large Radio Galaxy Sample}

Figure~4 shows the observed distribution for a heterogeneous 
collection of radio galaxies from Ledlow \& Owen (1996), compared to
a representative Monte Carlo simulation for a sample of 200 radio galaxies 
from a complete survey to a flux limit of 0.1~Jy at 1.4~GHz, 
with no redshift limit and $h=2$.
Our simulation nicely reproduces the almost uniform coverage of the 
$P-L$ plane in the region $-25<M_R<-21$~mag and $23<\log P<28$~W/Hz, 
with maximum concentration around the center of this region. 

This sample is not statistically complete, so there is no point in
doing a formal statistical comparison with our simulated data. 
We present it as an illustrative case, to show that radio sources 
cluster in this area of the $P-L$ plane. There are four reasons
for this:
\begin{itemize}
\item
Brighter galaxies do not exist in the volume sampled because of 
the steepness of the optical luminosity function at the bright end.

\item
No radio galaxies are observed fainter than $M_R\sim -22$~mag because the 
function $N(L)$ decreases rapidly at low optical luminosities.

\item
The minimum observed radio power is set by the radio flux limit of the 
survey and by the limited volume over which these faint objects can be 
discovered. Decreasing the flux limit results in extending toward lower 
radio power the region populated by observed radio galaxies.

\item
The maximum observed radio power is set by the rapidly 
decreasing probability of having luminous radio sources, i.e., by
the steepness of the bright end of the radio luminosity function.
\end{itemize}

\subsection{Comparison with a Volume-Limited Radio Galaxy Sample}

A well-defined sample of radio galaxies has been studied by Fasano
Falomo \& Scarpa (1996), with final results presented by Govoni \ea
(2000). The original sample includes all radio galaxies with a certain 
area of the sky, in the redshift range
$0.01<z<0.12$ down to a flux limit, at 2.7~GHz, of 2~Jy 
for part of the sample and
0.25~Jy for the rest. The average Cousin R magnitudes are
$<M>=-24.00$~mag for the whole sample, and $<M_{FR~I}> = -24.13$~mag and 
$<M_{FR~II}> = -23.62$~mag, for the FR~I and FR~II sub-classes, respectively. 

In Figure~5 we show the results from a Monte Carlo simulation 
for the appropriate flux limits and redshift range.
The agreement is quite good: in more than 80\% of a set of 100 simulations,
a two-dimensional K-S test indicates the two samples are statistically
indistinguishable (i.e., probability they are from different 
populations $<5$\%).

The Govoni et al. sample includes both FR~I and FR~II, so we can study the
cumulative distributions separately. 
The simulated data follows exactly the observed distribution
for each type, with the same offset in optical luminosity 
of $\sim 0.5$~mag (Fig.~6).
Similar results were found in all 100 Monte Carlo simulations we ran.

The apparent difference in host galaxy magnitude is essentially a
selection effect, due to the steepness of the radio and optical
luminosity functions combined with the diagonal separation between
FR~I and II in the radio-optical luminosity  plane.  As described 
in \S~4.1, the radio galaxy density in the $P-L$ plane is
modulated by the radio luminosity function (more sources at low
power); in the optical plane, it is centrally peaked because the
optical luminosity function (more sources at faint magnitudes) is
multiplied by the probability of having a radio source (which
increases as $L^2$).  Given the diagonal division, then, FR~II can
have relatively modest radio powers (so there are relatively more of
them) when found in fainter galaxies, shifting their mean optical
magnitude toward the left.  In contrast, FR~I are more common in
brighter galaxies, where they can span a larger range of radio power;
thus their mean optical magnitude shifts to the right.

\subsection{Comparison to a Cluster-Based Radio Galaxy Sample}

Ledlow \& Owen (1996) presented radio and optical
luminosities for a sample of 188 radio sources found in 
Abell clusters; the sample is complete to $z=0.09$ for
radio flux greater than 0.01~Jy at 1.4~GHz. 
In Figure~7 we compare the observed data to 
a representative simulation for a survey reaching the 
same radio flux and including the whole volume out to 
the same redshift limit. The exponent of the probability
function was fixed at $h=2$. 

In both real and simulated data 
there are no very bright radio sources because of the 
limited volume surveyed. In particular, essentially all 
simulated radio galaxies lie below the transition line and 
should be FR~I, as observed by Ledlow \& Owen (1996).

However, a formal two-dimension K-S test gives a probability of 
only $1$\% that the two sample are from the same population, and
the probability remains below few percent in all the simulations we did.
Basically, our simulations produce brighter galaxies than 
observed in the real data. 
Better agreement between observed and simulated data can be 
obtained setting
$h=1.8$, which is well within the uncertainty. In this case, the two samples 
are statistically indistinguishable 20\% of the times.

It is not clear if this result reflects the failure of one or more
of our basic assumptions or if some systematic effects are playing a role. 
For instance, the selection function for Abell clusters may
weight the volume sampled by Owen \& Ledlow 
in a way we do not take into account.
In addition, radio sources in clusters may not be a representative
subset of galaxies everywhere, just as galaxies themselves
can be different in clusters and in the field.
For example, galaxies in clusters may have undergone
more mergers, on average, so that their luminosity distribution 
deviates from the optical LF we are using. 
It is not clear whether these effects could make
the observed properties of this radio galaxy sample different
from our simulation and from the other radio galaxy samples.

\section{Discussion and Conclusions}

We have shown that the optical luminosity function of radio
galaxies is consistent with their having been drawn from 
a parent population of normal elliptical galaxies, weighted
by the function $S(L) \propto L^h$. 

This conclusion depends only on the optical luminosity function
of ellipticals and the slope of the weighting function.
This implies that any elliptical galaxy has a finite probability
of forming a radio source, with more luminous galaxies being
more likely to be radio-loud than faint galaxies.

Comparing the derived optical luminosity distribution 
for radio galaxies with a combined sample from the literature
we constrain the slope to be $h=2\pm0.4$.
This result depends somewhat on the adopted luminosity function 
for elliptical galaxies (see \S~3). 

We then used Monte Carlo simulations to show that under quite general
assumptions --- that radio and optical powers are independent and that
the separation of FR~I and FR~II galaxies is a diagonal line in the
radio-optical plane --- simulated samples of radio galaxies match well
the properties of observed samples. The agreement with 
data Govoni \ea (2000) is excellent, while
the match with the cluster-based sample of Ledlow \& Owen (1996) is 
qualitatively similar but quantitatively less good, perhaps because
of the cluster selection.

This implies a continuity of radio and optical properties 
from normal elliptical to radio galaxies. 
That is, there is no clear separation between
radio-loud and radio-quiet ellipticals, at any fiducial optical
luminosity or radio power. Moreover, our result is consistent with a
picture in which FR~I and FR~II radio sources are hosted by galaxies
extracted from the same parent population. No intrinsic differences 
are necessary to explain the well known difference of $\sim 0.5$ mag in
optical magnitude between the two classes of radio galaxies. 
This is due to the transition
region being an increasing function of the optical luminosity,
which is probably due to the way the plasma propagates through the
interstellar medium (Bicknell 1995) and does not imply any peculiarity
in the host galaxy. So we conclude that the apparent difference in 
observed optical magnitude is just a consequence of the probability density
contours in the $P-L$ plane.

The main conclusion from this work is that available data 
are consistent with the hypothesis that normal elliptical galaxies are
the parent population of elliptical radio galaxies. 
The physical interpretation for this continuity of elliptical galaxy properties
across all radio powers is that all ellipticals have a central black hole
and therefore have the potential to generate radio sources.
This hypothesis is supported by increasing empirical evidence for the
existence of massive black holes in the cores of nearby elliptical
galaxies (Kormendy \& Richstone 1995; Richstone et al. 1998; 
van der Marel 1999; Macchetto \ea 1999).
Once the radio source is created, its power should depend on the
accretion rate and the black-hole mass. Though the latter is found to be
proportional to the galaxy total mass (Ferrarese \& Merritt 2000,
Gebhardt et al. 2000), the former should depend on the availability
of gas and stage of development of the accretion activity. It is not
too surprising, therefore, that the radio power is largely independent
of the magnitude of the host galaxy. 

The fact that very often radio
galaxies exhibit a disturbed optical morphological or are found in
interacting systems is not an exclusive properties of radio galaxies,
but rather a common behavior of giant ellipticals, and indeed many
interacting galaxies or galaxies with peculiar morphology are known
not to be radio sources. Thus we conclude that the present results
explain nicely an important characteristic of radio-loud and
radio-quiet ellipticals, namely that they can not be distinguished solely 
based on optical observations (e.g., Ledlow \& Owen 1995; Govoni et al. 2000).

\acknowledgements
We thanks the referee for helpful comments that greatly improved
the paper, and P. Padovani, R. Falomo, and L. Maraschi for
help during the development of this work. 
Support for this work was also provided by NASA
through grants GO-06363.01-95A and GO-07893.01-96A
from the Space Telescope Science Institute,
which is operated by AURA, Inc., under NASA contract NAS~5-26555.

\newpage

\begin{figure}
\psfig{file=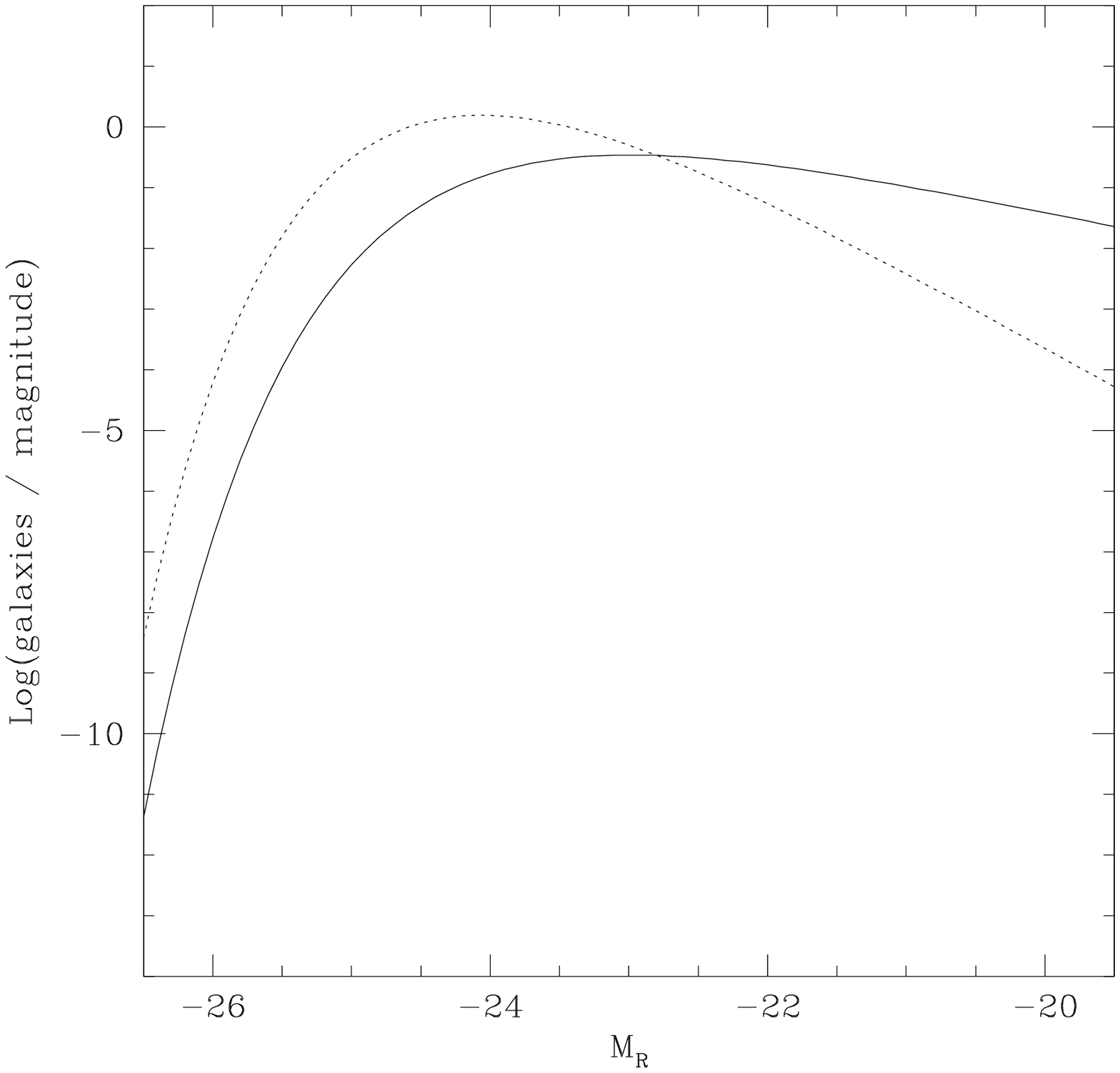,width=13cm}
\caption{
Observed optical luminosity function for elliptical galaxies 
({\it solid line}, arbitrary units),
a Schechter function with $M^*_R=-22.8$~mag and $\alpha = +0.2$ 
(Loveday \ea 1992). The fraction of small galaxies is very
large and diverges if $\alpha < -1$. 
{\it Dotted line:} Predicted optical luminosity function for radio galaxies,
obtained by multiplying the above by $S(L)=L^2$ 
(the probability for being a radio source).
This luminosity function is peaked around $M_R\sim -24$~mag, and the
fraction of small galaxies is dramatically reduced, as observed. 
The strong dependence of the probability $S$ on $L$ diminishes
the impact of any uncertainty in $\alpha$, because for any reasonable
value, the sum $\alpha+2$ is positive, making the exact shape of
the faint end of the LF unimportant.
}
\end{figure}

\begin{figure}
\psfig{file=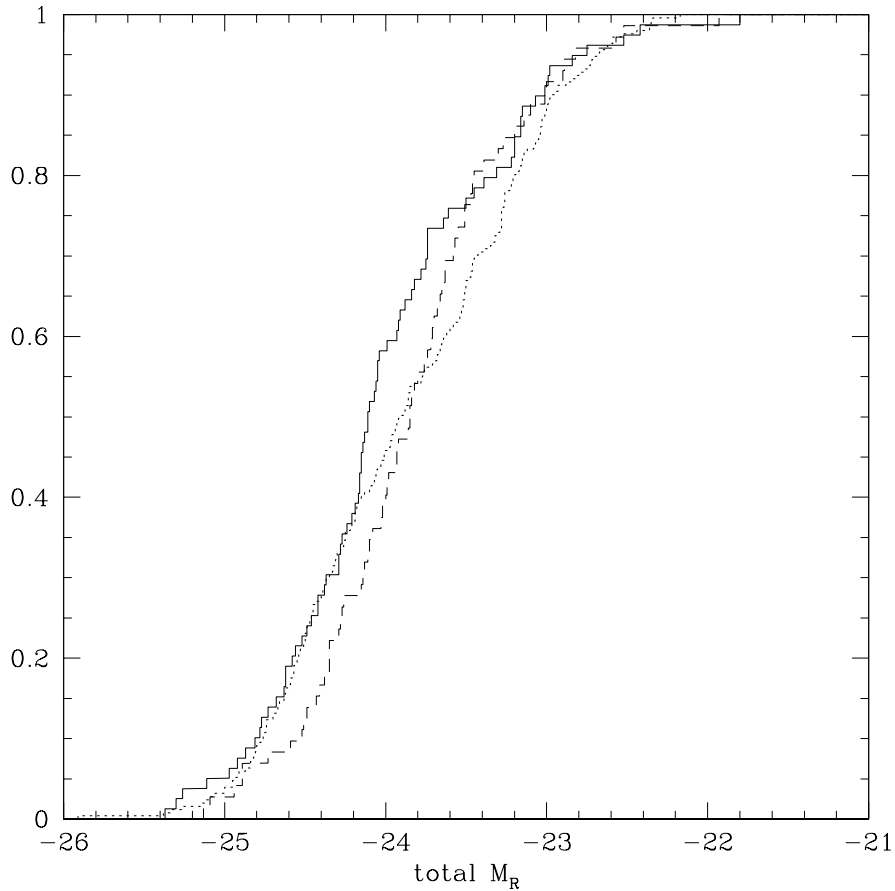,width=13cm}
\caption{
Comparison of absolute optical magnitudes for radio galaxies from
three different samples. All data were originally corrected for
interstellar extinction and K-correction, and have been transformed to
our adopted cosmology $H_0=50$ km/s/Mpc and $q_0=0$. 
{\it Solid line:} data from Govoni et al. (2000); 
these are total magnitudes in Cousin
R band obtained fitting a de Vaucouleurs law to the observed
luminosity profile. {\it Dotted line:} data from Ledlow \& Owen (1996), 
transformed from $M_{24.5}$ (i.e., aperture magnitudes extended
to $\mu_R= 24.5$ mag/asec$^2$) to total magnitudes applying a fixed
correction of $-0.2$ magnitudes. 
{\it Dashed line:} data from Smith \& Heckman (1989),
transformed from V to R band assuming
(V-R)=0.8, and further adding $-0.15$ to pass from $M_{25}$ to total
magnitudes. The distributions agree very generally but, given 
the large number of objects in each sample, the probability that 
these data sets are from the same population is only $\sim 5$\%
according to a K-S test.
}
\end{figure}

\begin{figure}
\psfig{file=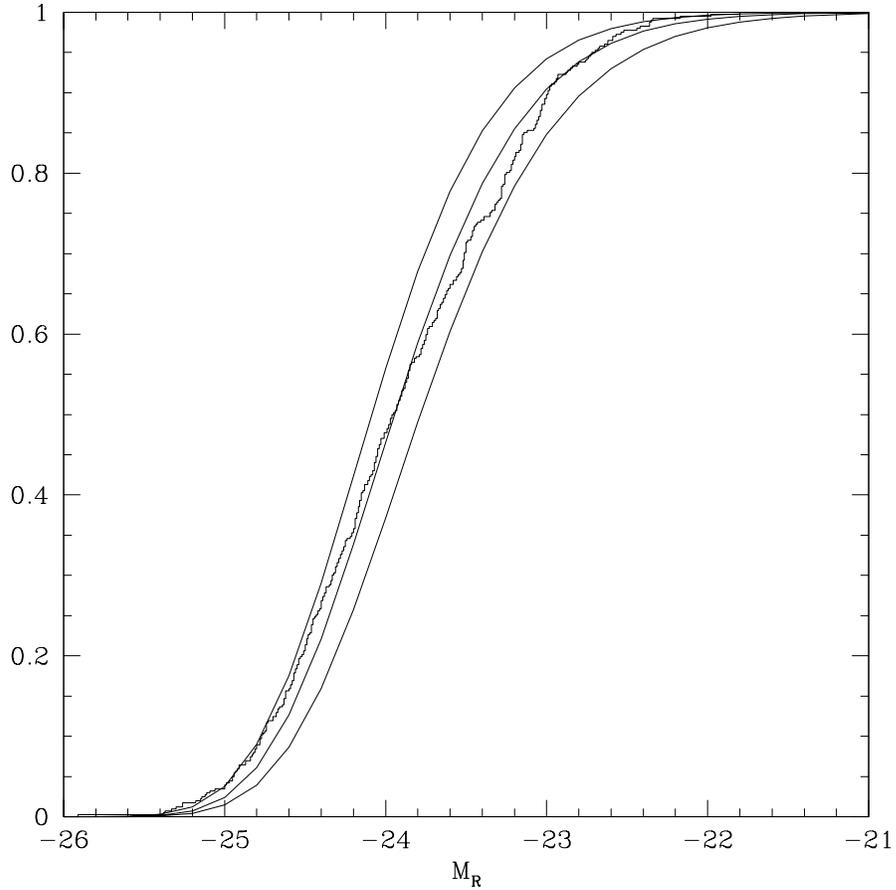,width=13cm}
\caption{
The combined data set of total optical luminosities for 
the three radio galaxy samples plotted in Fig.~2 ({\it histogram}). 
This combined sample is used to determine the exponent 
of the probability function 
linking radio-quiet to radio-loud elliptical galaxies. 
The abscissa gives the luminosity in units of $M=10^{-0.4(L/L^*}+M^*$. 
Superposed to the observed data are the expected cumulative 
distributions for radio galaxies ({\it curves}), 
an incomplete gamma function $\gamma(1+\alpha+h , L/L^*)$, 
for $h=$2.4, 2.0, 1.6, from left to right, respectively.
}
\end{figure}

\newpage
\begin{figure}
\centerline{
\psfig{file=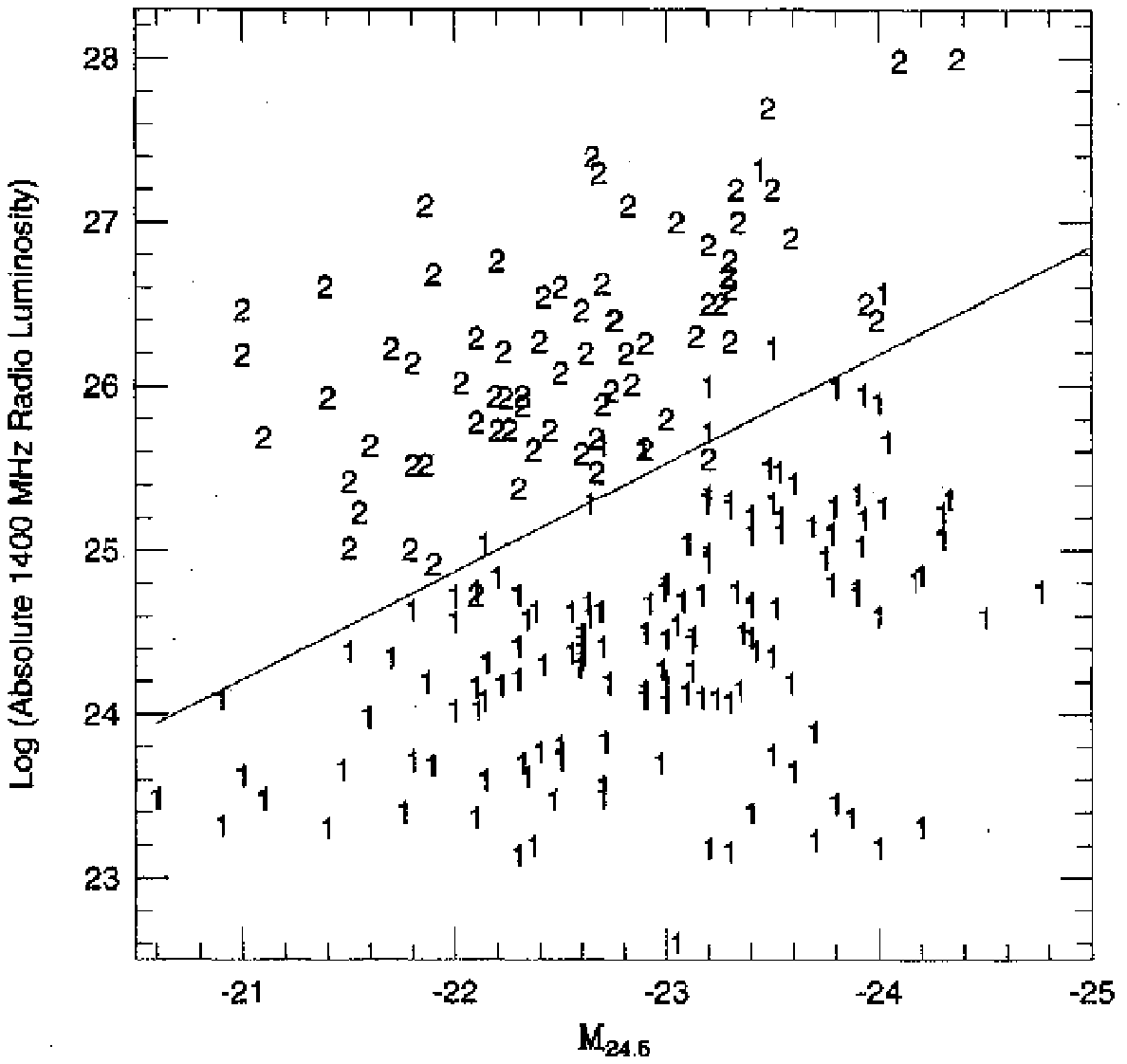,width=0.5\linewidth}
\psfig{file=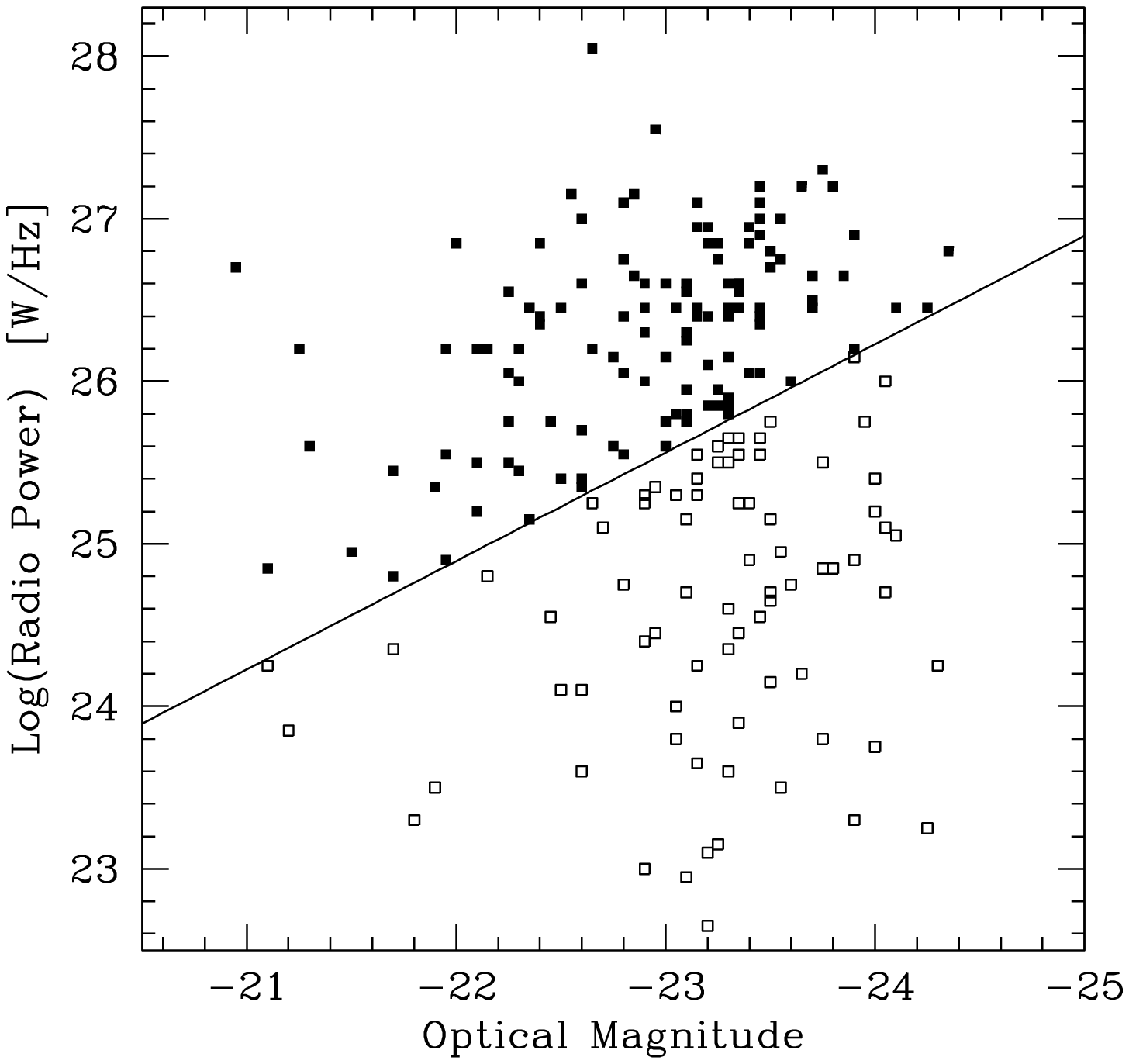,width=0.485\linewidth}
}
\caption{Radio power versus optical magnitude for radio galaxies
from multiple radio surveys.
{\it Left:} Observed distribution of FR~I ({\it symbol 1}) and
FR~II ({\it symbol 2}) radio galaxies in the radio-optical plane,
from the heterogeneous sample of Ledlow \& Owen (1996; their Fig.~1). 
The solid line separating FR~I from FR~II is the one originally 
drawn by Ledlow \& Owen.
{\bf Right:} Representative Monte Carlo simulation for
a complete flux-limited survey to 0.1~Jy at 1.4~GHz, no redshift limit,
and $h=2$. The simulation nicely reproduces the almost
uniform coverage of the plane in the region $-25<M_R<-21$~mag and
$23<Log(P)<28$~W/Hz, with maximum concentration around the center of
this region. Solid squares represent FR~II, open squares FR~I, defined
entirely by their position with respect to the solid line (same as
in left panel). For consistency with Ledlow \& Owen (1996), 
this figure was computed with $H_0=75$ km/s/Mpc.
}
\end{figure}

\newpage
\begin{figure}
\centerline{
\psfig{file=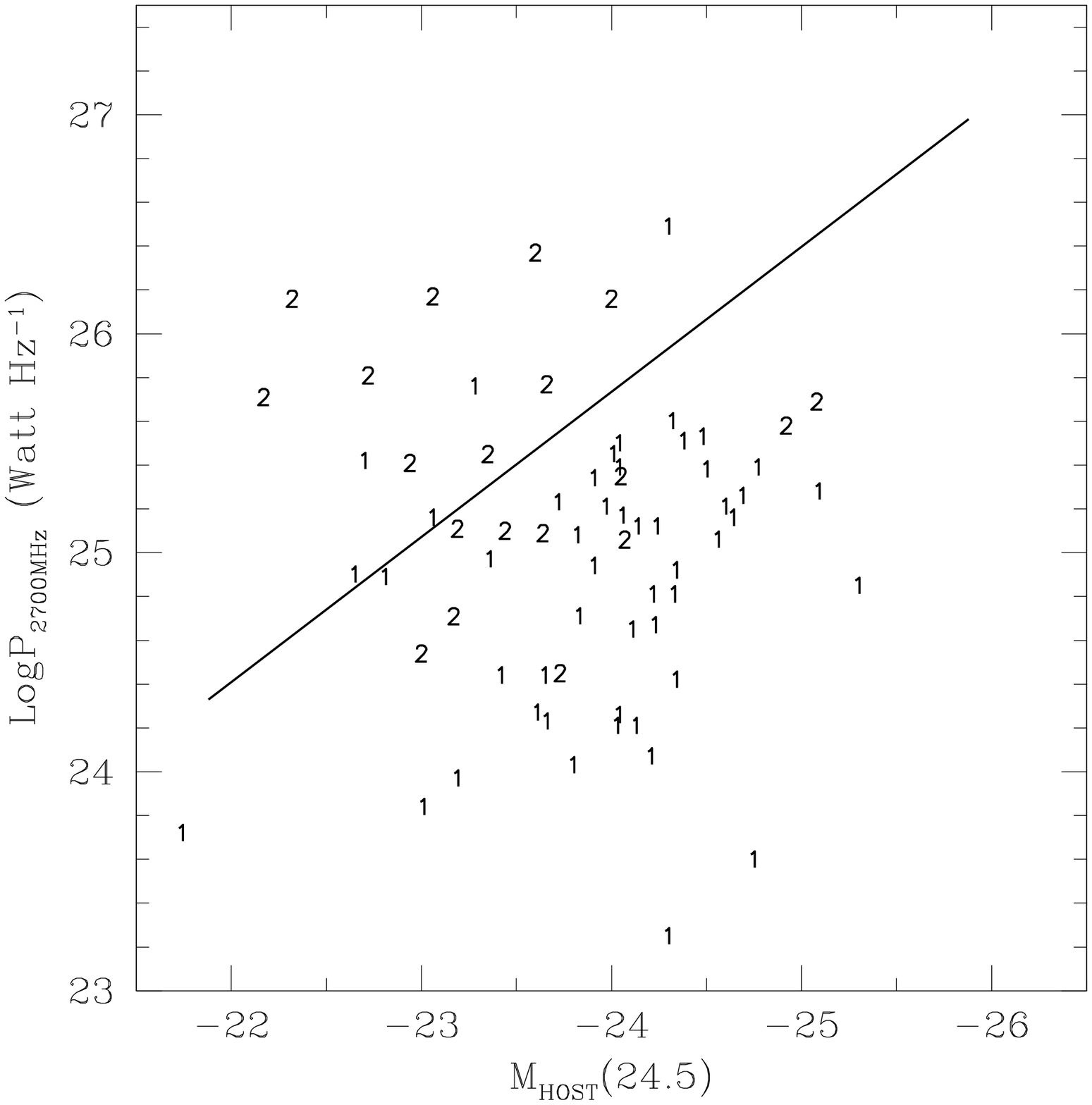,width=0.40\linewidth}
\psfig{file=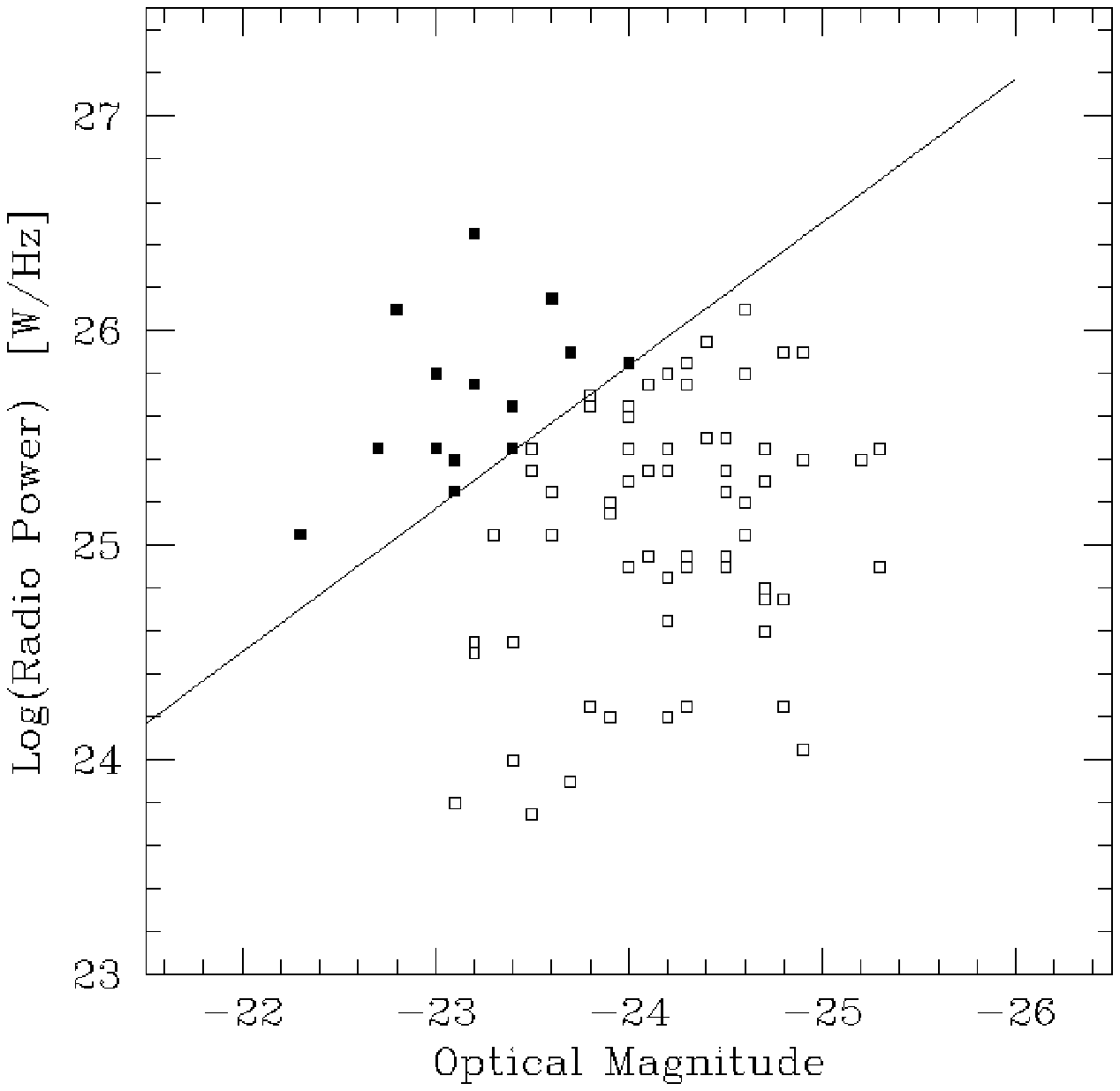,width=0.40\linewidth}
}
\caption{Radio power versus optical magnitude for radio galaxies
from a well-defined volume-limited survey.
{\bf Left:} Distribution for a sample of radio galaxies 
in the redshift range $0.01<z<0.12$, 
to a flux limit, at 2.7~GHz, of 2~Jy 
for part of the sample and 0.25~Jy for the rest,
from Fasano, Falomo \& Scarpa (1996) and Govoni \ea (2000).
{\bf Right:}
Monte Carlo simulation matched to the Govoni \ea selection criteria.
The agreement is excellent in both
the distribution of sources in the radio-optical luminosity plane
and the relative populations of FR~I (open squares)
and FR~II (solid squares). 
For consistency, this figure is computed with $H_0=50$ km/s/Mpc.
}
\end{figure}

\newpage
\begin{figure}
\psfig{file=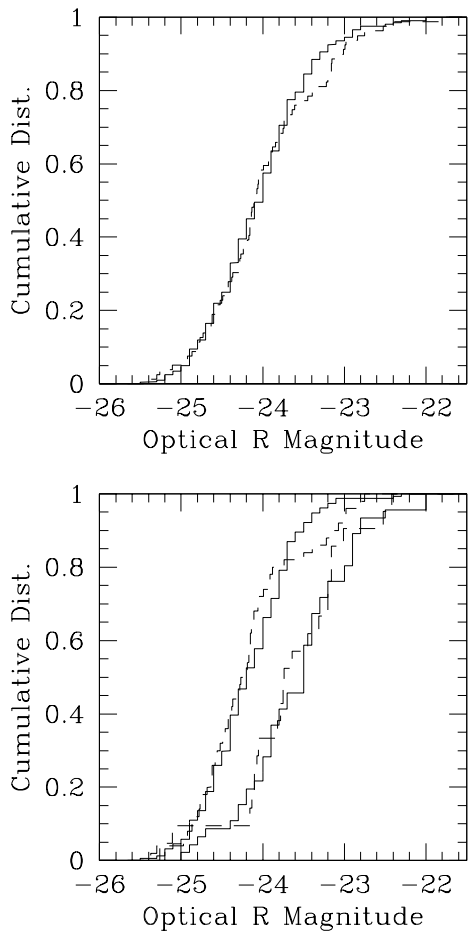,height=16cm}
\caption{
Cumulative distributions of R-band absolute magnitudes for
the radio galaxy data and simulations in Fig.~5. 
{\it Solid line:} simulated data; 
{\it dashed line:} observed data.
{\bf Left:} Cumulative distribution for the full data set of
Govoni et al. (2000). The probability the two sample are from the same
population is 60\% according to the K-S test. {\bf Right:} Separate
cumulative distributions for FR~I ({\it left}) and FR~II ({\it
right}). The simulated and observed data agree very well (K-S
probability that samples are from the same population is 30\% and 90\%
for FR~I and FR~II, respectively), indicating that the observed
difference in average optical luminosity between FR~I and FR~II is
essentially a selection effect. It arises because of the interplay
between radio and optical luminosity functions, modulated by the
diagonal division between FR~I and II. That is, powerful radio sources
are rare, and so those found tend to be in the more numerous,
lower-optical-luminosity galaxies, whereas FR~I radio sources are more
common and can be found in more rare, higher-optical-luminosity
galaxies.
}
\end{figure}

\newpage
\begin{figure}
\centerline{
\psfig{file=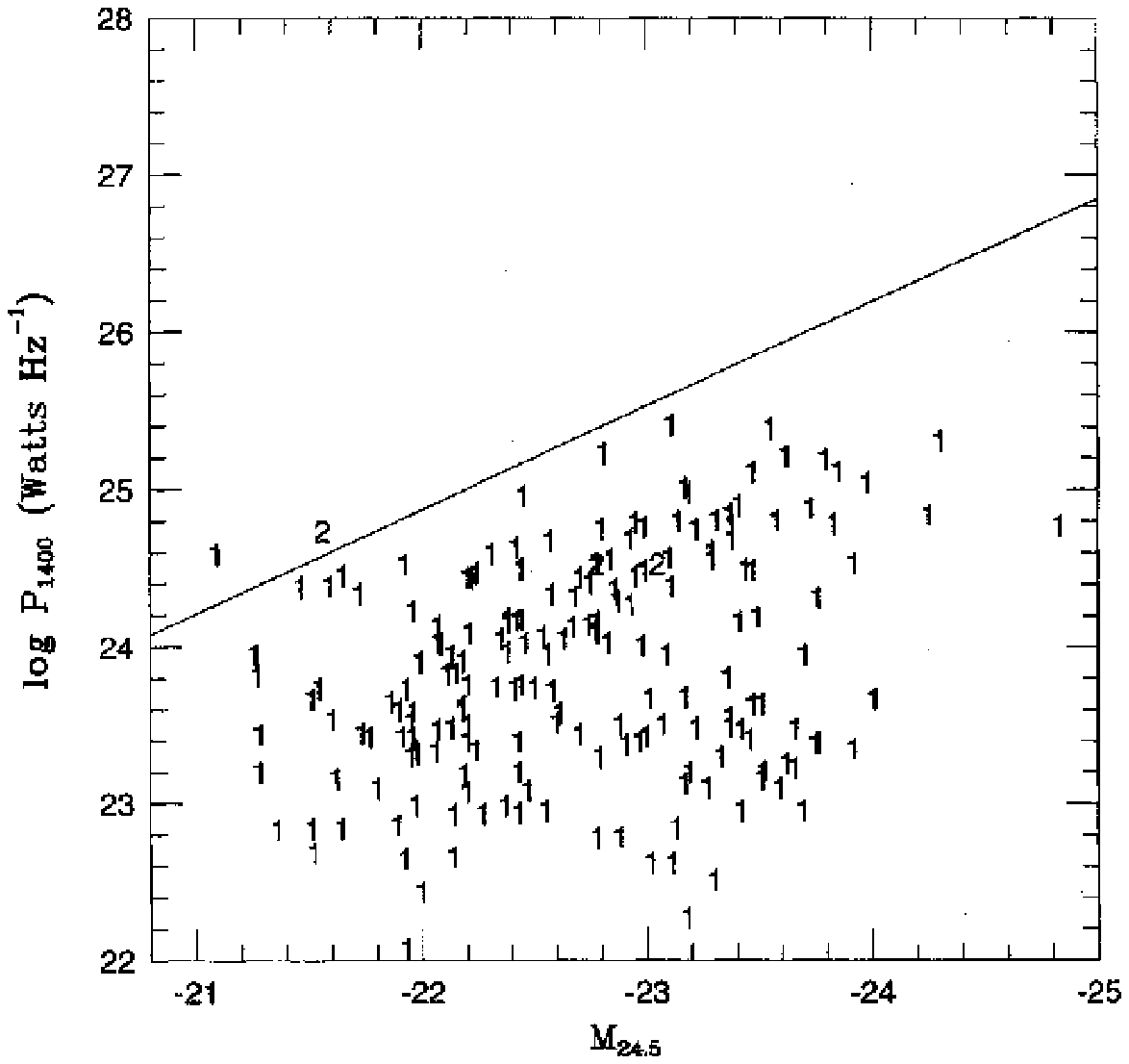,width=0.48\linewidth}
\psfig{file=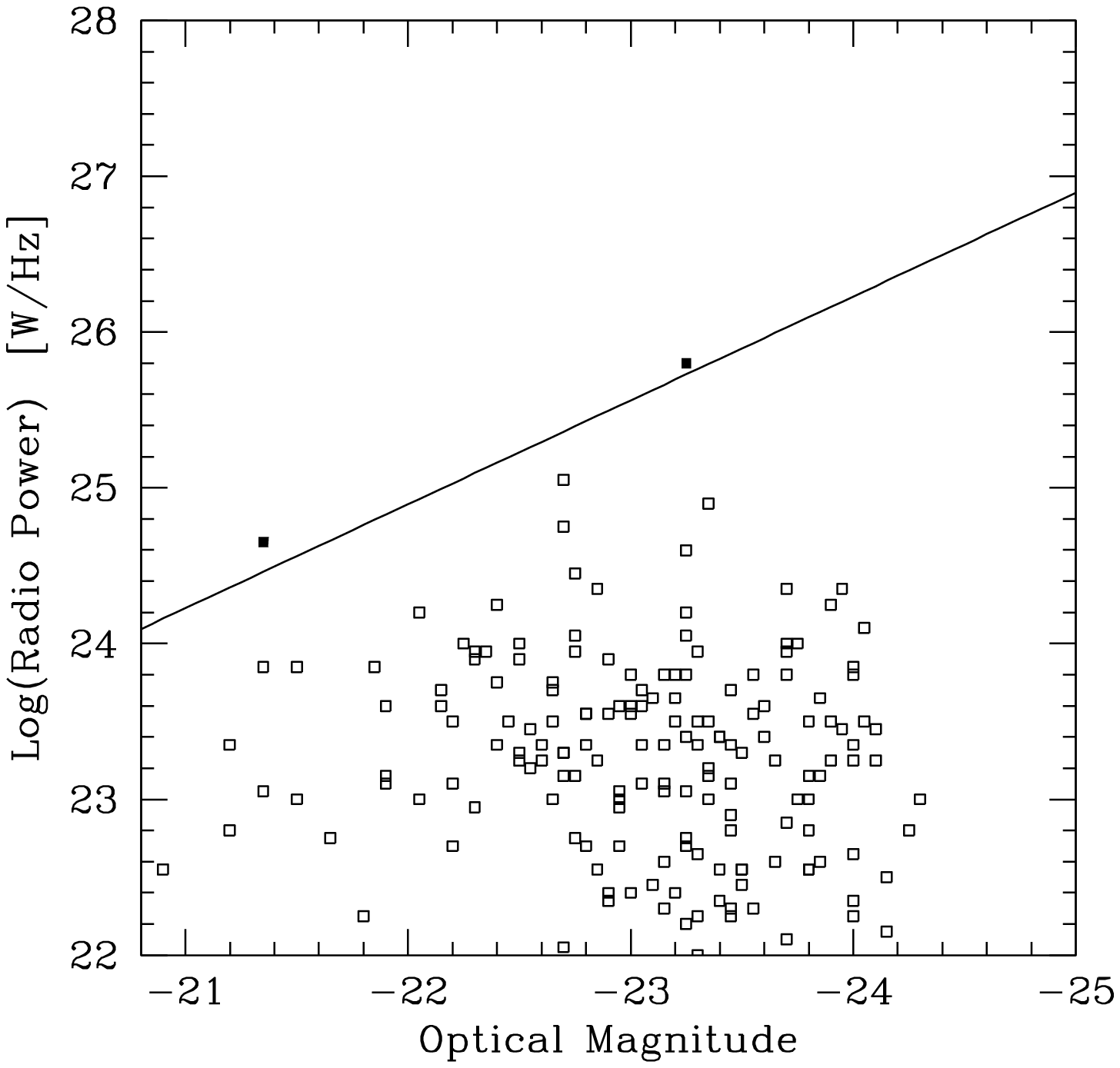,width=0.48\linewidth}
}
\caption{Radio power versus optical magnitude for radio galaxies
from a well-defined cluster-based survey.
{\bf Left:} Observed radio and optical luminosities for a
sample of 188 radio sources in the cores of Abell clusters, 
complete to redshift $z=0.09$ to a limiting radio
flux of 0.01~Jy at 1.4~GHz, from Ledlow \& Owen (1996). 
{\bf Right:} Distribution derived from a Monte Carlo simulation matched
to the same selection criteria. 
The two agree qualitatively:
there are no very powerful radio sources because of the small volume
surveyed (basically all sources are below the
transition line and should be FR~I, as indeed 
observed by Ledlow \& Owen).
However, a K-S test implies the distributions differ significantly,
as does the Ledlow \& Owen sample from other radio galaxy samples,
which may be related to the cluster selection criterion.
For consistency with Ledlow \& Owen (1996), 
this figure was computed with $H_0=75$ km/s/Mpc.
}
\end{figure}

\end{document}